\documentclass[
  journal=largetwo,
  manuscript=article-type,
  year=2020,
  volume=37,
]{cup-journal}

\usepackage{amsmath}
\usepackage[nopatch]{microtype}
\usepackage{booktabs}
\usepackage{amssymb}

\newcommand{\cm}{\mbox{\,cm}}

\newcommand{\degK}{\,\mbox{K}}

\newcommand{\grad}{^{\mbox{\small o}}}

\newcommand{\kms}{\mbox{\,km s}^{-1}}
\newcommand{\kpc}{\mbox{\,kpc}}
\newcommand{\EJ}{E_{\rm J}}

\newcommand\Msun{\mbox{M}_\odot}

\newcommand{\Myr}{\mbox{\,Myr}}
\newcommand{\Gyr}{\mbox{\,Gyr}}
\newcommand{\pc}{\mbox{\,pc}}
\newcommand{\pcc}{\mbox{\,cm}^{-3}}

\title{Gas Dynamics in the Central Molecular Zone and its connection with the Galactic Bar}

\author{Leonardo Chaves-Velasquez}
\affiliation{Instituto de Radioastronomía y Astrofísica, Universidad Nacional Autónoma de México, Apdo. Postal 3-72, Morelia, Michoac\'an 58089, M\'exico}
\alsoaffiliation{Astronomical Observatory, University of Nariño, Sede VIIS, Avenida Panamericana, Pasto, Nariño, Colombia}
\email[Leonardo Chaves-Velasquez]{leonardochaves83@gmail.com}

\author{Gilberto C. Gómez}
\affiliation{Instituto de Radioastronomía y Astrofísica, Universidad Nacional Autónoma de México, Apdo. Postal 3-72, Morelia, Michoac\'an 58089, M\'exico}

\author{Ángeles Pérez-Villegas}
\affiliation{Instituto de Astronomía, Universidad Nacional Autónoma de México, A. P. 106, C.P. 22800, Ensenada, B. C., México}


\begin{document}

\begin{abstract}
The innermost region of the Milky Way harbors the central molecular zone (CMZ). This region contains a large amount of molecular gas but a poor star formation rate considering the densities achieved by the gas in this region. We used the arepo code to perform a hydrodynamic and star formation simulation of the Galaxy, where a Ferrers bar was adiabatically introduced. During the stage of bar imposition, the bar strength excites density waves close to the inner Lindblad resonance guiding material toward the inner Galaxy, driving the formation of a ring that we qualitatively associate with the CMZ. During the simulation, we
identified that the ring passes three main phases, namely: formation, instability, and quasi-stationary stages. During the whole evolution, and particularly in the quasi-stationary stage, we observe that the ring is associated with the x2 family of periodic orbits. Additionally, we found that most of the star formation occurs during the ring formation stage, while it drastically decreases in the instability stage. Finally, we found that when the gas has settled in a stable x2 orbit,
the star formation takes place mostly after the dense gas passes the apocenter, triggering the conveyor-belt mechanism described in previous studies.
\end{abstract}


\section{Introduction}
The innermost region of the Milky Way (MW; $\rm R_{gal}<200 \pc$) harbours the central molecular zone (CMZ; \citealt{1996ARA&A..34..645M}). This region contains a large amount of molecular gas whose volumetric density is at least two orders of magnitude higher than that found in the Galactic disc \citep{2007A&A...467..611F,2011ApJ...735L..33M,2013MNRAS.429..987L,2014MNRAS.440.3370K,2016A&A...586A..50G,2017ApJ...835..263B}. Observational studies confirm the existence of dense cores of cold gas with low star formation rate (SFR;\citealt{Yusef-Zadeh_2009,2012A&A...548A.120I,2013MNRAS.429..987L,2017MNRAS.469.2263B}). The gas in this region is subjected to extreme physical conditions compared with the star formation peak in galaxies in the universe's early stages \citep{2010Natur.463...65C,2003ApJ...599.1116C,2016A&A...586A..50G,2016A&A...595A..94I,2017ApJ...850...77K,2017MNRAS.470.1462L,2018ApJ...868....7M,2019A&A...630A..74M}. 

The CMZ hosts massive young clusters (MYCs), such as Arches and Quintuplet (these clusters achieve masses $\gtrsim 10^{4} M_{\odot}$; 
\citealt{1999ApJ...525..750F}). \citet{2015MNRAS.449..715W} compare the distribution of the mass as a function of the ratio with the possible progenitor clouds of these clusters, finding a conflict between these two quantities, concluding that the conveyor-belt mechanism is more suitable in this process than the monolithic collapse for the formation of the MYC.

\begin{figure*}
    \centering
    \includegraphics[width=\textwidth]{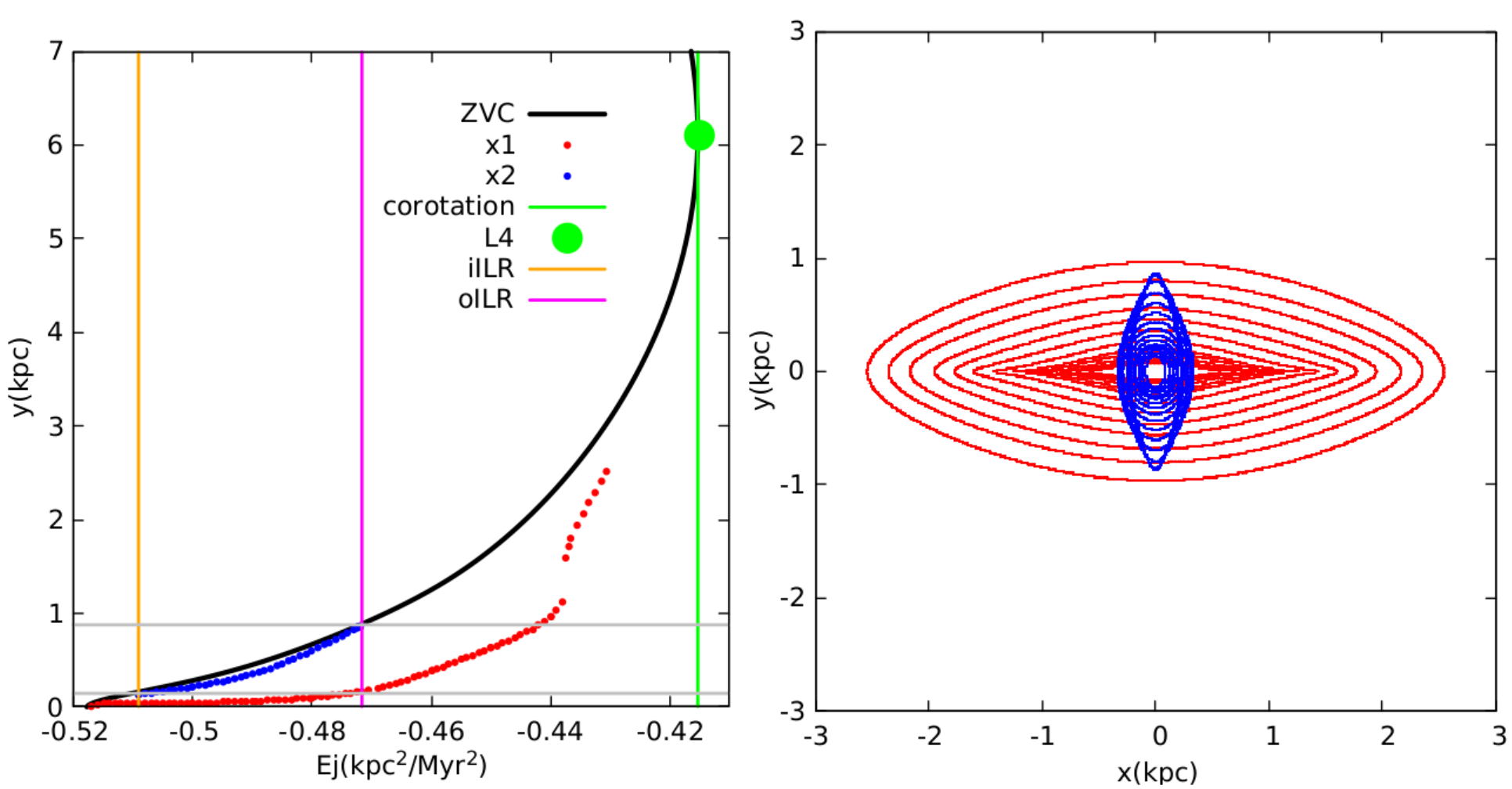}
    \caption{({\it Left}): Jacobi energy ($\EJ$) vs. the intersection with the $y$-axis for periodic orbits of the x1 ({\it red dots}) and the x2 ({\it blue dots}) families in the autonomous model. The black curve corresponds to the zero velocity curve. The green, orange, and magenta vertical lines show the position of the corotation radius, the iILR, and the oILR, respectively. The horizontal grey lines show the extension of the x2 region. 
    ({\it Right}): A sample of members of the x1  and the x2 orbit families (red and blue lines, respectively).}
    \label{fig:figura_001}
\end{figure*}

An open question that remains in discussion is the morphology of the CMZ. As first proposed by \citet{1991MNRAS.252..210B}, the gas that follows orbits corresponding to the x1 family switches to the x2-type \footnote{The x1 orbital family is composed of orbits aligned with bar and gives it the orbital support. On the other hand, the x2 orbital family is perpendicular to the bar and confined within the inner Lindblad resonance. See right panel of Figure  \ref{fig:figura_001}.} of orbits forming an inner disk-like structure. In that sense, the existence of these two families of periodic orbits must be conditioned to the presence of a non-axisymmetric structure, like a Galactic bar.

The Galactic bar in the inner region of the MW has been clearly established,
both through photometric and dynamical studies. The first evidence of the existence of a Galactic bar was obtained by \citet{1991ApJ...379..631B} by analyzing the $2.4 \,\mu m$ emission from the Galactic center. Later on,  \citet{1995ApJ...445..716D} used near-infrared data from COBE DIRBE to characterize the morphology of the Galactic bulge, and they found that the bulge resembles a barred-shaped structure with its end near the first quadrant of the Galactic plane. Other important observational works support the fact that the MW is a barred galaxy \citep{1997MNRAS.288..365B, 2013Msngr.154...54W}.  From the dynamical point of view, the first Galactic models that included a bar were proposed by \citet{mulder}.
Other works that include the dynamical effects of a Galactic bar in order to fit  structures observed in different data sets  include \citet{1999MNRAS.304..512E}, \citet{2003MNRAS.340..949B}, \citet{2008A&A...489..115R}, \citet{2015MNRAS.449.2421S}, \citet{2015MNRAS.454.1818S}, \citet{2016ApJ...824...13L}, among others.

Recently, there has been theoretical work focused on exploring the dynamical origin of the CMZ.
\citet{Kruijssen+15} reported a best fit open orbit (a rosette), integrated in an axi-symmetric background potential, that in principle adjusts the positions of the largest molecular clouds in this region. However, \citet{2020MNRAS.499.4455T} claim that there is no inconsistency between the open orbit and scenarios that associate the CMZ to an x2 orbit since the molecular clouds could follow quasi-periodic trajectories around a given x2 periodic orbit. On the other hand, \citet{2018MNRAS.481....2S} predicts the size of stable rings based on the fact that regions of reverse shear are favorable to form stable rings if the sound speed is low.

The three-dimensional (3D) structure of the CMZ is still controversial. In the literature, we can find kinematic studies of longitude-velocity diagrams in order to unravel the actual 3D morphology of this region \citep{1995PASJ...47..527S,1998ApJS..118..455O,1999ApJS..120....1T,Kruijssen+15,2017MNRAS.470.1982S,2016MNRAS.457.2675H,2019PASJ...71S..19T,2022MNRAS.516..907S}. In \citet{2022MNRAS.516..907S} by making a detail comparison between CO and HI maps and LV diagrams, found that CMZ is a  plateau distribution of molecular gas distributed in a ring-like structure that traces a double-loop shape, as seen projected onto plane of the sky.   
\citet{2016MNRAS.457.2675H} studied the kinematics of dense gas in the region containing the CMZ and shows the distribution of gas velocities, as traced by HNCO, in $l$-$b$ space.
\citet{2022MNRAS.516..907S} traced the CMZ using CO and HI lines, concluding that it lies in a range of $l\sim \pm 2\grad$  and $b\sim \pm 0.5\grad$.

Despite the studies of gas and stellar dynamics associated with the Galactic bar, its connection with the CMZ has not been explored in detail. In principle, the self-interacting gas that follows x1 orbits shocks in the regions of high curvature and falls to the x2 region following the so-called dust channels \citep{2015MNRAS.449.2421S,2020MNRAS.499.4455T,Hatchfield_2021}. Depending on the nature of the potential, there are regions (close to the 4:1 resonance) in which members of the x1 family develop big loops. In these cases, the dust  channels tend to avoid this regions and form straight-line shocks \citep{2022Univ....8..290P}. \citet{2000A&A...358...45P}, by using SPH simulations of strong bar models, concluded that the inflow of the gas via the dust channels increases as the sound speed increases. On the other hand, the increase of the inflow is stronger when the bar is imposed in an abrupt manner. Recently, \citet{Sormani+24} proposed that the bar potential excites strong trailing density waves that generate a gap close to the ILR, removing angular momentum and transporting the gas inwards. In this scenario, the mechanism to form the CMZ would originate around the ILR as opposed to the gas dynamics at bar scales.

 Regarding star formation, despite the high densities achieved in the CMZ, the star formation rate (SFR) found in this region is low compared with the outer parts of the Galaxy. Observations of the dust ridge at $870\,\mu\text{m}$ confirm that the region contains one of the most massive reservoirs of molecular gas found in the Galaxy, in which the temperatures achieved by the densest clouds are low, suggesting a very early stage of star formation \citep{2012A&A...548A.120I}. In \citet{2013MNRAS.429..987L} by taking into account the low activity of star formation in the CMZ, the authors inferred a relevant physical mechanism that suppresses star formation. This mechanism relates with an additional term or threshold in the star formation relations which takes into account the turbulent energy that would suppress gravitational collapse. On the other hand, it is worth to consider that the gas in the central region of the Galaxy is subjected to extreme physical conditions, including tidal forces that could reduce the critical mass for gravitational collapse \citep{2023MNRAS.524.4614Z}.

 In this work we developed a hydrodynamic simulation in {\sc arepo} in which we imposed an external axisymmetric potential composed by a halo, a disc, and a bulge. The bulge, adiabatically transfers part of its mass to a Ferrers bar during the first $667 \Myr$ of the simulation. During this imposition, we observed an internal ringed structure. This ring passes three main phases, namely: formation, instability, and a quasi-stationary state. We analyzed the density maps in the three phases and created plane-of-the-sky projections to observe the evolution of the structure in $l-b$ diagrams. Additionally, we studied the gas flow induced by the bar during the phases of the ring. We also studied the star formation and its relation with the geometry of the ring. This paper is organized as follows: In section \ref{simulation} we explained the set-up of the simulation, we introduce the dynamical and geometric parameters of the external potential, and the process of imposition of the bar. In section \ref{cmzx2} we  discuss the x2 region given by the autonomous model by computing periodic orbits of the x1 and x2 family. Also in section \ref{cmzx2} we discuss the main phases of the ring, namely formation, instability, and quasi-stationary stages. In section \ref{sec:flux} we study the gas flow at different radii. In section \ref{sec:sfr} we analyze the star formation activity in the ring and we relate the SFR with the geometry of the ring. Finally, in section \ref{conclu} we present the main conclusions of our work.

 \section{Simulation}
\label{simulation}
We performed the simulation using the moving-mesh code {\sc arepo} \citep{2010MNRAS.401..791S,2020ApJS..248...32W}. We imposed the axi-symmetric potential from \citet{1991RMxAA..22..255A}, with the parameters proposed by \citet{2013A&A...549A.137I}. We summarize these parameters in Table \ref{table1}. This model comprises a disc, a dark halo, and a bulge. In addition to the axisymmetric part we impose a rigid ellipsoidal Ferrers' bar that rotates clockwise with a pattern speed of $\varOmega_{p}=40 \kms \kpc^{-1}$. The pattern speed of the bar is one of the most important dynamic parameters since it determines the location of the resonances. The value used in this work is consistent with recent observational studies that have constrained this parameter. For instance in \citet{2017MNRAS.465.1621P}, with a combination of VVV, UKIDSS, 2MASS, BRAVA, OGLE and ARGOS surveys, the estimated value for the pattern speed estimated to be $\varOmega_{p}=39\pm 3.5 \kms \kpc^{-1}$. Other works in which the estimated pattern speed is close to the value used here include \citet{2002A&A...384..112L, 2015MNRAS.454.1818S, Bland-Hawthorn2016, 2019MNRAS.488.4552S, 2020MNRAS.497.5024S}.
\begin{table}
\centering
\caption{Parameters of the Galactic Model.}
\label{table1}
\begin{tabular}{ |p{3cm}||p{3cm}|  }
 \hline
 \multicolumn{2}{|c|}{Axisymmetric Components} \\
 \hline
Parameter& Value\\
 \hline
 $M_{\rm b}$    & $9.48\times10^{11}\text{M}_{\odot}$\\
 $M_{\rm d}$    &$6.63\times10^{12}\text{M}_{\odot}$\\
 $M_{\rm h}$    &$2.36\times10^{12}\text{M}_{\odot}$\\
 $b_{\rm b}$    & $0.23 \kpc$\\
 $a_{\rm d}$    &$4.22 \kpc$\\
 $b_{\rm d}$    &$0.292 \kpc$\\
 $a_{\rm h}$    &$2.562 \kpc$\\
 $\Lambda  $    &$200 \kpc$\\
 $\gamma   $    &$2$\\
 \hline
 \multicolumn{2}{|c|}{Galactic bar} \\
 \hline
 Parameter& Value\\
 \hline
 $M_{\rm bar}$   & $6.32\times10^{11}\text{M}_{\odot}$\\
 $a_{\rm bar}$   & $3.5 \kpc$\\
 $b_{\rm bar}$   & $1.4 \kpc$\\
 $c_{\rm bar}$   & $1 \kpc$\\
 \hline
 \multicolumn{2}{@{}p{8cm}@{}}{
    $M_{\rm b}$: Bulge mass,
    $M_{\rm d}$: disc mass,
    $M_{\rm h}$: halo mass,
    $b_{\rm b}$: buldge radial scale parameter,
    $a_{\rm d}$: disc radial scale parameter,
    $b_{\rm d}$: disc vertical scale parameter,
    $a_{\rm h}$: halo radial scale parameter,
    $\Lambda$: halo cut off parameter, 
    $\gamma$: free parameter introduced to avoid singularities at the origin,
    $M_{\rm b}$: bar mass, 
    $a_{\rm bar}$: bar semi-major axis,
    $b_{\rm bar}$: bar semi-minor axis, and 
    $c_{\rm bar}$: bar vertical semi axis.
}\\
 
\end{tabular}
\end{table}

In order to set the initial density distribution of the gaseous Galactic disc, we defined a radius $R_{\rm in} = 3 \kpc$ and set the midplane density constant for $R < R_{\rm in}$ and with an exponential profile with a $8 \kpc$ scale for $R>R_{\rm in}$. The gas density at $R = 8 \kpc$ was set as $1 \pcc$. In the vertical direction, the density is defined following hydrostatic equilibrium with the background axisymmetric potential. We distributed $5 \times 10^6$ gas cells in the simulation at random positions following this density profile leading to an initial resolution of approximately $2\times 10^4 \Msun$ per cell. Since {\sc arepo} allows for mass flow between cells, this resolution does not remain uniform during the evolution.

The gas is initially at a uniform temperature of $8 \times 10^3 \degK$, but it is then allowed to thermally evolve. We used the cooling function proposed by \citet{KoyamaInutsuka02}\footnote{The typographical corrections outlined in \citet{Vazquez+07} were used.}, using the exponential cooling model described in \citet{Vazquez+07}. This function induces a thermally bistable behavior in the gas with an unstable regime for densities in the approximate range of 1 through $10 \pcc$.

To avoid transient effects, the bar is imposed adiabatically by transferring part of the mass of the bulge to the bar \citep{2000AJ....119..800D} during the first 667 Myr, so that the bar's final mass is $\sim 79 \%$ of the initial bulge's mass.
This approach has been adopted in other works including
\citet{kim1,seo1,armillota1,seo2,armillota2,li1}. After this time, 
the background gravitational potential remains constant, although the gas' self-gravity is calculated throughout the simulation. 

However, if we assume that the hydrodynamic forces and gas self-gravity are negligible compared to the general orbital dynamics, the gas kinematics may  be approximated to the ballistic model of test particles under the potential as applied to the simulation at that time. Under this approximation, the purpose of retaining a fraction of the bulge's mass is to allow for a suitable x2 orbital region that the gas could follow. Our simulation is stopped after $5 \Gyr$ of evolution.

\begin{figure}
    \centering
    \includegraphics[width=\columnwidth]
    {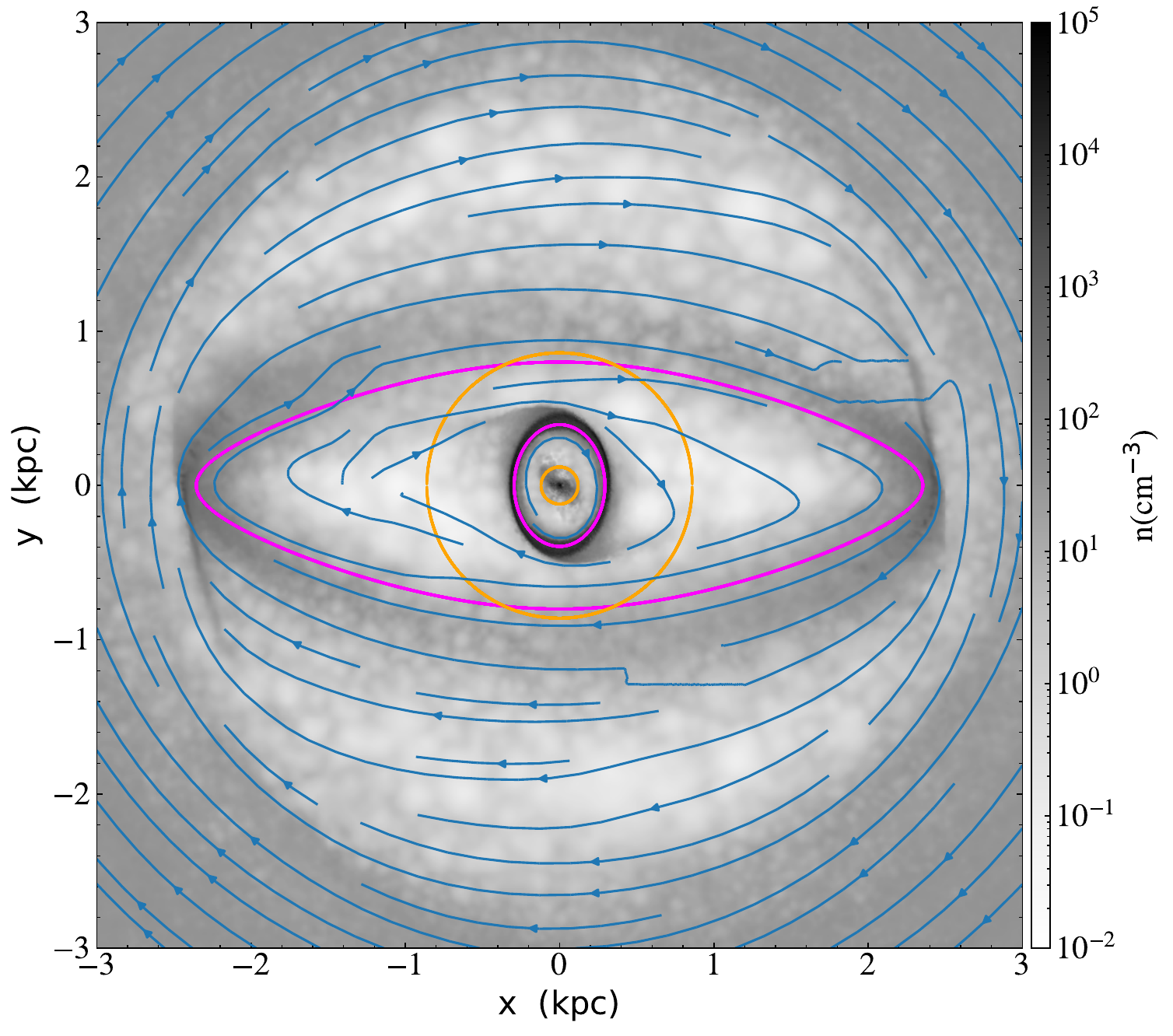}
    \caption{Density map at $t=2000 \Myr$. The magenta lines correspond to two orbits from the x1 (outer line, $\EJ= -0.444 \kpc^{2}/\Myr^{2}$) and x2 family (inner line, $\EJ=-0.486 \kpc^{2}/\Myr^{2}$). The orange circles denote the positions of the iILR and the oILR.
    }
    \label{fig:den_or_001}
\end{figure}

\section{CMZ: The x2 region}
\label{cmzx2}

The morphology of the CMZ has been associated with a ring-like structure of molecular gas that is confined to 
the inner region of the MW Galaxy \citep[$R<250 \pc$;][]{2012MNRAS.424..666Z,2020ApJ...902....9O,2020MNRAS.497.5024S}. 
Similar structures have been observed in barred external galaxies and are associated with the gas flow that originates from the galactic bar tips. As the orbital dynamics pushes the gas to follow trajectories similar to x1 periodic orbits \citep{GomezEtal2013}, it shocks and forms the observed dust channels \citep{1992MNRAS.259..345A}.
The CMZ is associated with x2 periodic orbits \citep{2003ApJ...582..723R,2015ApJ...806..150L,2018MNRAS.481....2S}.

The phase-space available for x2 orbits depends on the existence and location of the inner Lindblad resonance (ILR). The presence of a massive bulge might split the ILR into two: inner and outer (iILR and oILR, respectively; \citealt{1989A&ARv...1..261C,1992MNRAS.259..328A}). 
If this is the case, the x2 orbits will survive only between the iILR and oILR \citep{1982MNRAS.201..303V}, and so the gaseous ring of the CMZ could exist only in this region. In this work, we set the extension of the x2 region by retaining some fraction of the bulge contribution after the bar potential is imposed to the gaseous disc in the simulation (see section \ref{simulation}).

On the left panel of Figure \ref{fig:figura_001} we show the $\EJ$ vs $y$ diagram of the autonomous model. The zero-velocity curve (black line) and the energies corresponding to both inner Lindblad resonances existing in the  model (orange for the iILR, magenta for the oILR) are also shown.\footnote{Note that the location of these resonances are approximations from the autonomous model since the gas self-gravity should change their actual positions.} Corotation, defined as the position of the L4 Lagrangian point, is located at $y \approx 6.11\kpc$. The red dots denote the periodic orbits corresponding to members of the main x1 family and the blue points are members of the x2 family. We notice that the x2 region is limited by the iILR and the oILR, between $y \approx 0.12$ and $0.86\kpc$. On the right panel of Figure \ref{fig:figura_001} we show a sample of members of the x1 and x2 families of periodic orbits (in red and blue respectively). The x1 orbits support and shape the Galactic bar, while the x2 orbits support the internal ring, or in this case the CMZ.

Gas parcels try to follow periodic orbits (as long as the orbit does not cross itself) since the pressure forces are much smaller than the gravitational forces. Therefore, we should see gas structures associated to the bar (when the gas follows x1 orbits) or to the internal ring (when the gas follows x2 orbits).

In Figure \ref{fig:den_or_001} we show the gas density distribution after $2\Gyr$ of evolution. We observe a bar structure with an approximately $2.4 \kpc$ semi-major axis and an internal ring in the region between the two ILRs (defined with the two green circles). This ring structure appears to be generated by gas following x2 orbits. The internal magenta orbit corresponds to an x2 orbit that approximately matches the CMZ and the external magenta orbit corresponds to a member of the x1 family that matches the bar.

\begin{figure*}
\includegraphics[width=0.7\textwidth]{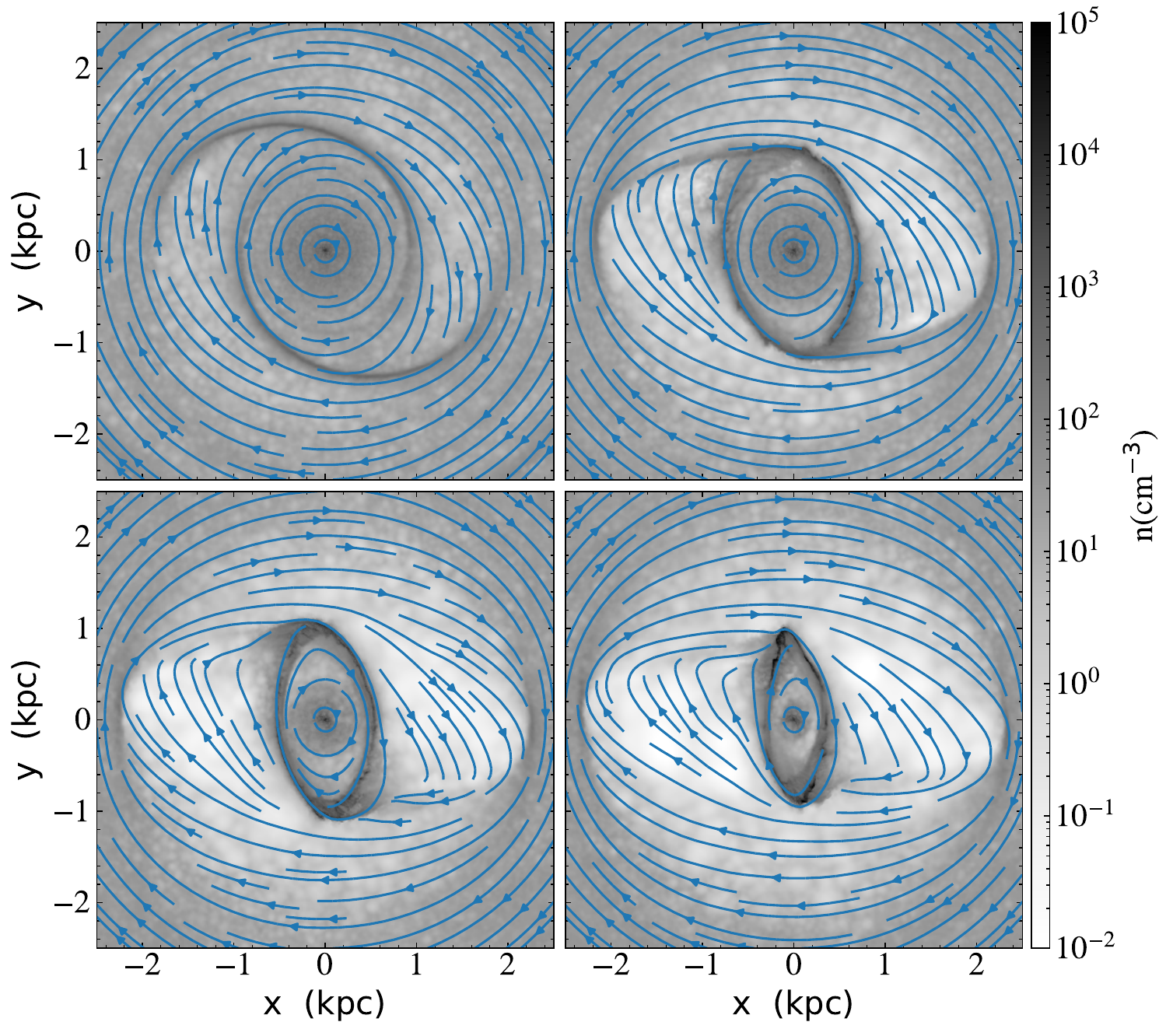}
    \caption{Midplane gaseous density at $t = 300$, $450$ ({\it top row}), $570$, and $650\Myr$ ({\it bottom row}) into the simulation. Blue streamlines trace the velocity field of the gas. During this time the model is evolving since the bar is growing at the expense of the bulge mass. It can be seen that the formation of the internal ring structure is due to the perturbation of trailing density waves from the bar.}
    \label{fig:fig_snaps_imp}
\end{figure*}

\begin{figure}
    \centering
    \includegraphics[width=\columnwidth]{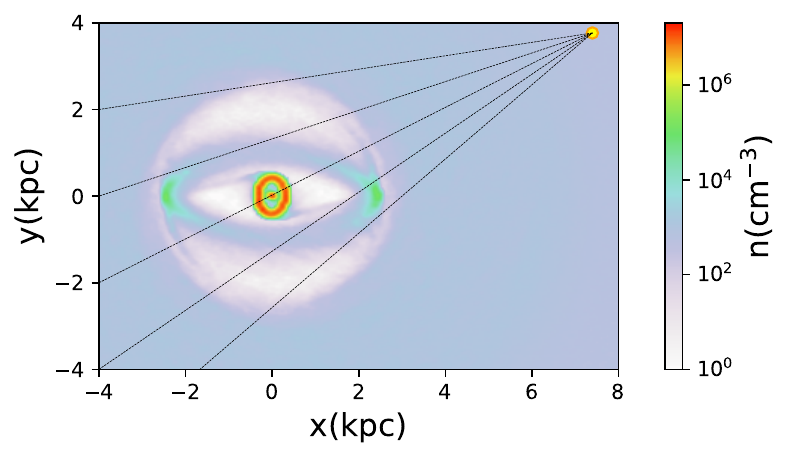}
    \caption{Position selected for the observer in the simulated $l-b$ visualizations. The semi-major axis of the Galactic bar forms an angle of $-27^{o}$ with respect to the line between the Galactic center and the Sun at $8.3 \kpc$ (yellow dot). Colours show the gaseous density distribution at $t=2 \Gyr$.}
    \label{fig:esquema_001}
\end{figure}

\begin{figure*}
    \centering
    \includegraphics[width=\textwidth]{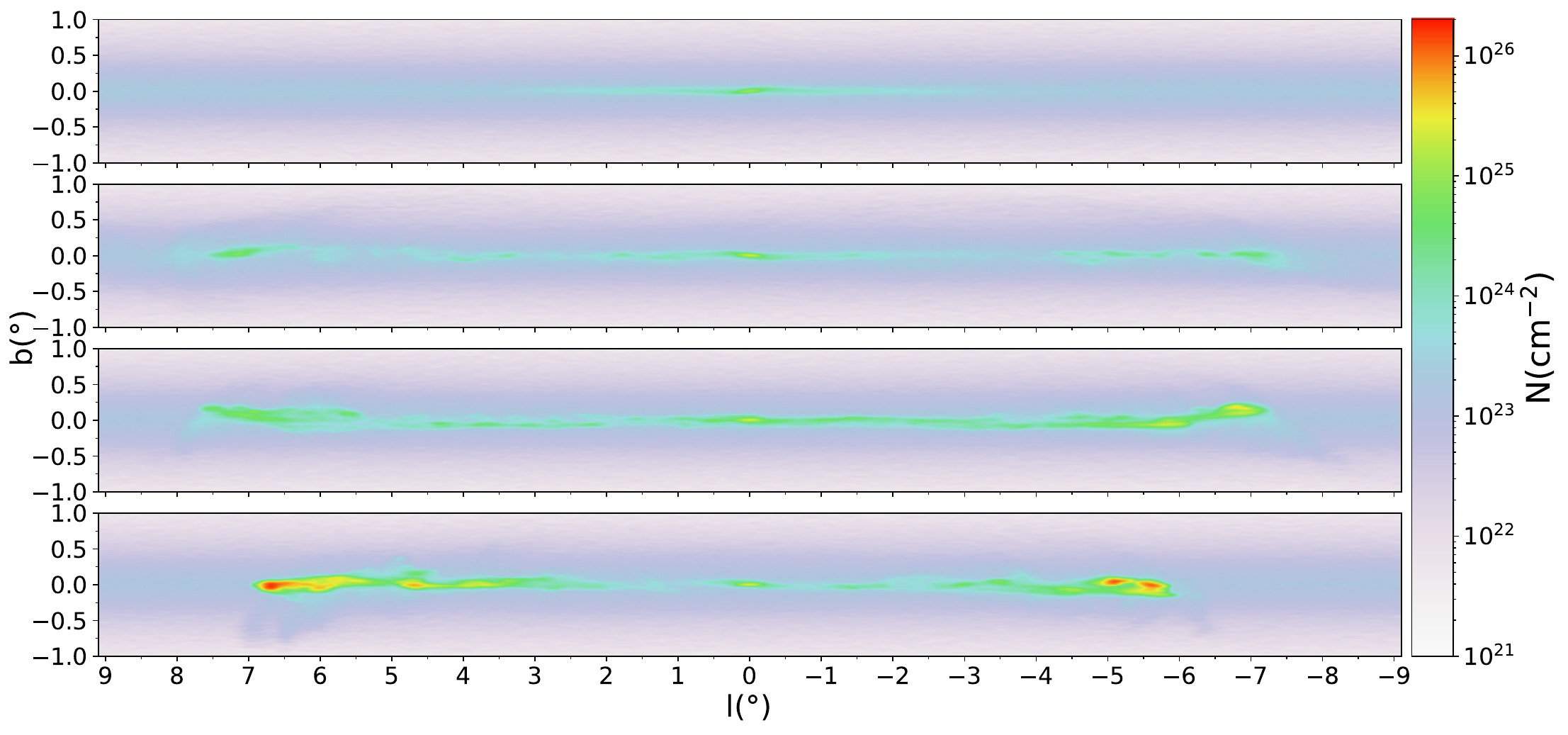}
    \caption{Views of the ring structure projected to the plane of the sky in Galactic coordinates ($l$-$b$) for the same times as in fig. \ref{fig:fig_snaps_imp}: down from the top panel, $t=300, 450, 570$ and $650 \Myr$.
    }
    \label{fig:l_b_001}
\end{figure*}

\begin{figure*}
    \centering
    \includegraphics[width=0.7\textwidth]
    {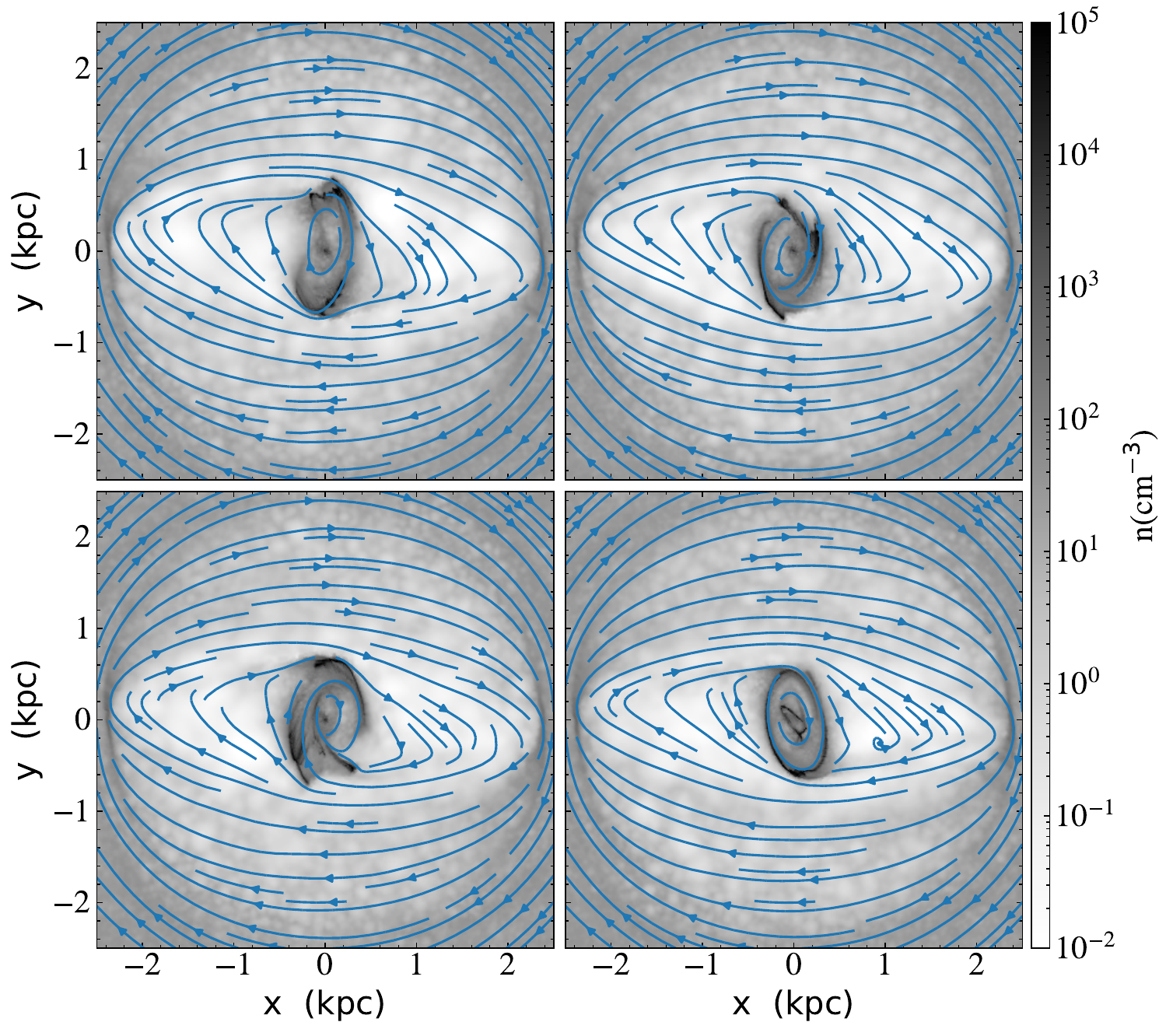}
    \caption{Gaseous density at $t = 730$ ({\it top Left}), $735$ ({\it top Right}), $745$ ({\it down left}), and $785\Myr$  ({\it down right}). During this lapse, the morphology of the ring manifests an irregular behavior which we attribute to self-gravity. 
    }
    \label{fig:snaps_chao}
\end{figure*}

\begin{figure*}
    \centering
    \includegraphics[width=\textwidth]{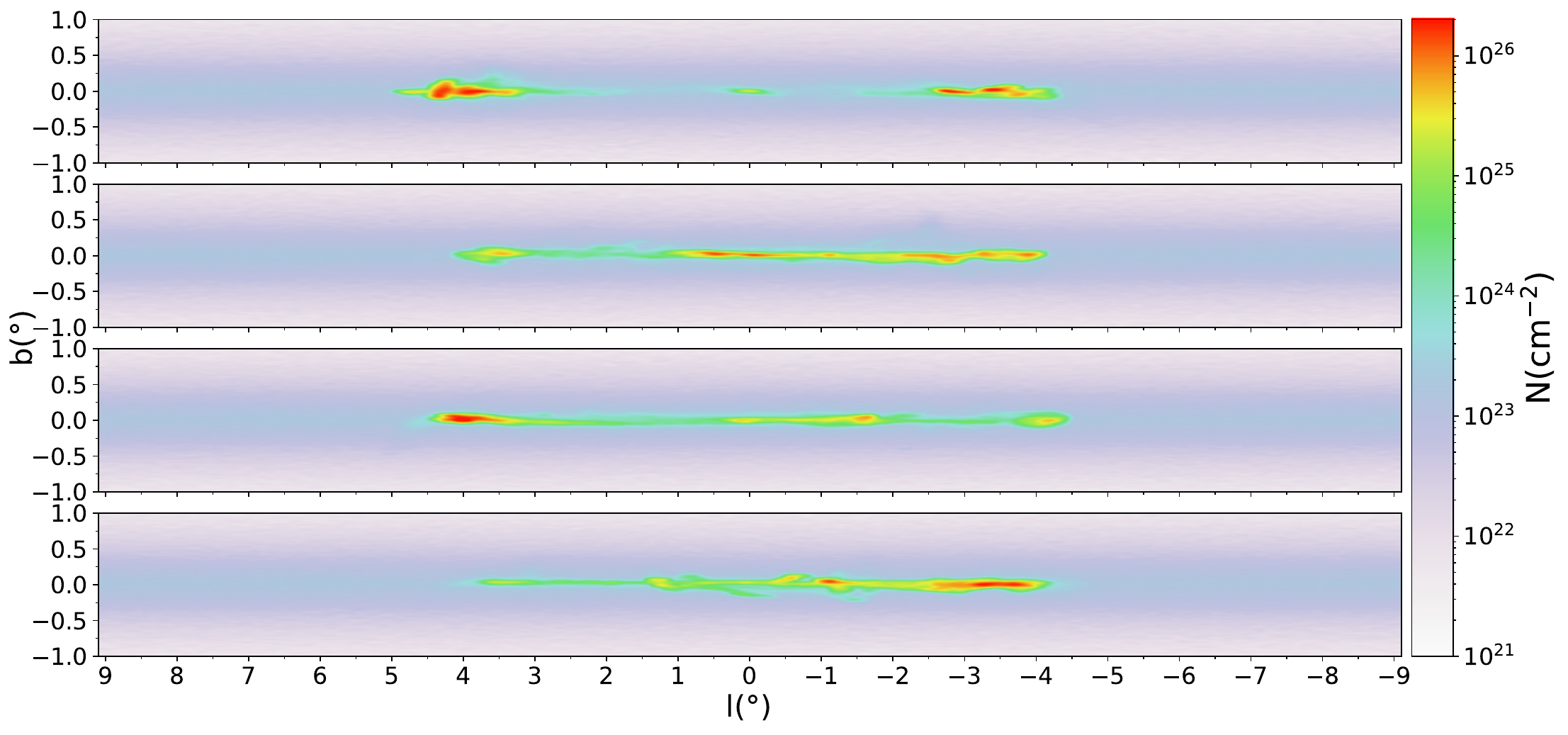}
    \caption{Evolution of the ring region projected on the sky. Down from the top, panels show the simulation at $t=730$, $735$, $745$ and $785 \Myr$.
    }
    \label{fig:l_b_002}
\end{figure*}

\subsection{Phases of the ring}
\label{sec:phases}

The internal ring formed in this simulation presents abrupt changes in morphology during the simulation. Three phases of evolution are distinguishable, which we identify as formation, instability, and quasi-stationary state. The main changes in morphology occur during the phases of formation and instability. After the instability phase, the ring achieves a stationary state in morphology and star formation. In this section, we discuss these phases.  

\subsubsection{Formation}
\label{sec:formation}

Figure \ref{fig:fig_snaps_imp} shows the evolution of the ring from $300$ through $650 \Myr$. Since this period roughly corresponds to the time during which the bar is imposed and the bulge mass diminished, the location of the resonances in the system slowly changes. 
At $t=300 \Myr$ spiral arms are formed in the gas within a $2 \kpc$ radius. According to \citet{Sormani+24}, these spirals correspond to density waves that have been excited by the bar potential in regions close to the oILR. In principle, these waves remove angular momentum from the gas and, consequently, the material is transported inside the oILR, forming the inner ring.
The streamlines show that the trajectory of the gas particles at this time is almost circular since the bar is still weak ($23\%$ of the bulge mass at this time). The spirals lead to an oval structure that becomes apparent at later times.  As the bulge becomes more massive than the bar, the model presents a strong contribution of the x2 orbital family and the gas begins to accumulate in this resonance. 

At $t=450 \Myr$, a well defined ringed structure appears with densities around $2\times10^{2} \pcc$ containing small regions above $10^{3} \pcc$. The velocity streamlines show distorted gas motion around the ring with over-densities at the ends of the remnants of the spiral. From these overdensities, dust channels leading to the tips of the (now oval) ring are clearly seen. These dust channels are the trajectories of the gas from the borders of the bar toward the ring.

At $t=570 \Myr$, the ring is more elongated and it has reduced its size, presumably since the increased mass of the bar has reduced the region where x2 orbits are allowed. Although the ring is smaller, its gas density is increased.
The dust channels are less visible in density but are still discernible as abrupt direction changes in the gas streamlines.

At $650 \Myr$, the imposition of the bar is almost complete. We observe regions near the apocenters of the ring with densities above $10^{4} \pcc$. Note that the ring eccentricity is higher than at previous times. Also, the ring is smaller because the region between the ILRs shrinks. At this time, the gas density within the bar potential reaches its lowest values.

To compare the kinematics and star formation in the ring to the observed CMZ, we placed an imaginary observer at $8.3 \kpc$ from the Galactic center, along a line inclined $-27\grad$ with respect to the bar's axis \citep{Wegg2013, Bland-Hawthorn2016}, as shown in Figure \ref{fig:esquema_001}.
Based in this geometry,
figure \ref{fig:l_b_001} shows the projections onto the plane of the sky corresponding to the same evolution times from Figure \ref{fig:fig_snaps_imp}. 
The projected ring slowly decreases its angular size in galactic longitude, finally reaching from $l \sim -6.5\grad$ to $\sim 7\grad$.

Between $t=667$ and $730 \Myr$, a little after the end of the imposition of the bar, the ring does not suffer significant morphological changes. However, the star formation reaches its highest activity in this period (see Sec. \ref{sec:sfr}). 

\subsubsection{Instability}
\label{sec:instability}

During the second stage of the ring evolution, the morphology of the ring changes rapidly for a period of almost $118 \Myr$. Figure \ref{fig:snaps_chao} shows density maps between $730$ and $785\Myr$ into the simulation. At the beginning of this period, the ring undergoes a strong disruption with large regions with densities above $10^{4}\pcc$. Since the external potential remains static at this stage, we attribute the high densities achieved to gravitational collapse within the ring. At later times, these two segments orbit the galactic center and the ring morphology approaches an x2 orbit again.

Figure \ref{fig:l_b_002} shows the plane-of-the-sky projection for this period. At $730 \Myr$, on the borders of the structure, we observe surface number densities above $10^{26} \cm^{-2}$ and the extension of the ring decreases even if the bar and bulge potentials remain unchanged and the x2-orbit region size is constant. Although high-density regions can be seen above or below the galactic midplane during the formation stage, during the instability stage these departures appear to be more organized, suggesting that the condensations might follow off-plane oscillating orbits.

\subsubsection{Quasi-stationary state}

\label{sec:stationary}

\begin{figure*}
    \centering
    \includegraphics[width=\textwidth]{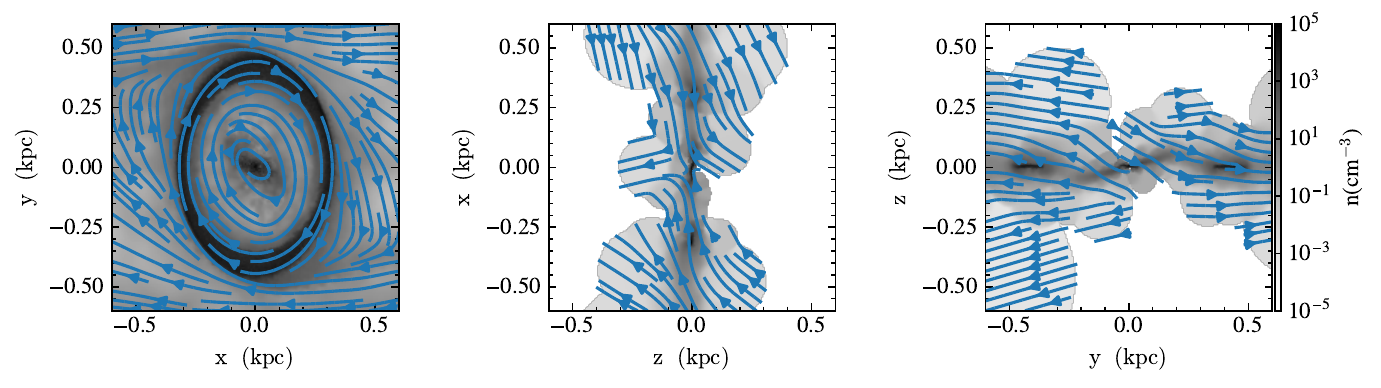}
    \caption{Density maps of the ring region at $t=2000\Myr$. Grayscale in the panels show cuts in the simulation box, while blue lines show velocity flowlines in the corresponding plane.}
    \label{fig:face_edge_001}
\end{figure*}

After the instability stage, the ring further reduces its size and enters a quasi-stationary state until the end of the simulation. Figure \ref{fig:face_edge_001} shows the three-dimensional structure of the ring region at $t=2 \Gyr$. Comparison of the $x-y$ plane ({\it left panel}) with Figure \ref{fig:den_or_001}, which also corresponds to this stage, shows that the ring closely follows an x2 orbit. Also, faint dust channels can be distinguished in Figure \ref{fig:den_or_001}.
\citet{2022MNRAS.516..907S} analyzed CO and HI line emission data cubes of the Galactic Center region and concluded that the CMZ seems to be embedded in the HI disc of radius $\approx 320 \pc$ and vertical scale height $\approx 70 \pc$. Their analysis suggests a ring-like structure that encloses an area from $l=-1\grad.1$ $(-157 \pc$) to $-1\grad.8$ $(257 \pc$). Our results at $2 \Gyr$ suggest a ring associated with the x2 orbital configuration of semi-major axis of $\approx 390 \pc$.
\begin{figure*}
    \centering
    \includegraphics[width=0.48\textwidth]{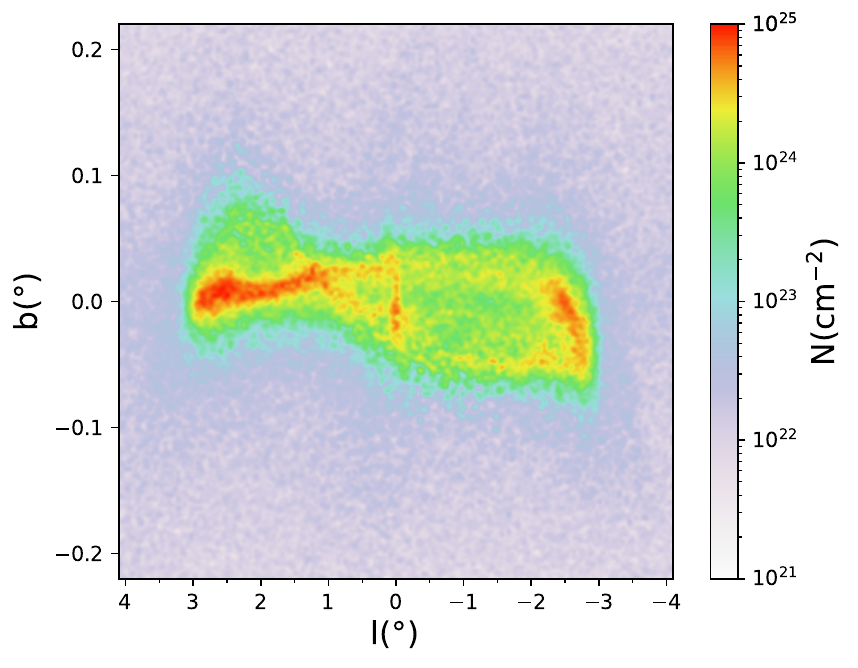}
    \hspace{0.02\textwidth}
    \includegraphics[width=0.48\textwidth]{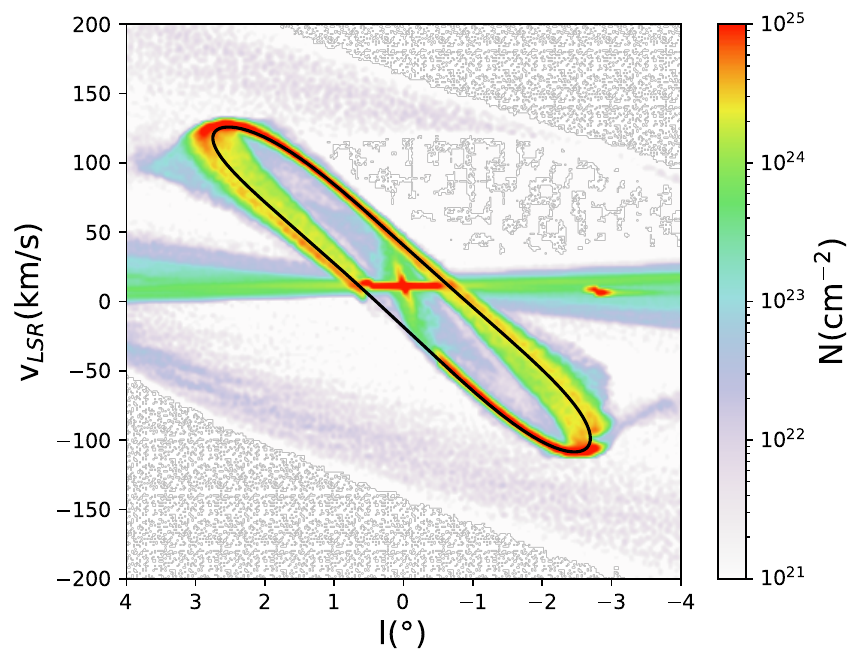}
    \caption{({\it Left}): Column density of the internal region at $t=2 \, \mathrm{Gyr}$. The ring structure lies mainly in the Galactic plane, with a width of $\sim 6^\circ$ in longitude. ({\it Right}): $l-V_{LSR}$ diagram at the internal region of the Galaxy at $t=2 \Gyr$. The black line is an x2 orbit that matches the CMZ.}
    \label{fig:l_b_400}
\end{figure*}
Density cuts along the $xz$ and $yz$ planes in Figure \ref{fig:face_edge_001} show that the ring is not completely planar but presents fluctuations characterized by excursions on the $z$ axis. The excursions on the vertical direction remain inside $|z|<0.1 \pc$.

The $l-b$ diagrams presented in Figure 4 of \citet{2022MNRAS.516..907S} show a warped ring traced by the Sgr B1, B2, C, D, and E molecular clouds. This trace indicates excursions of the ring out of the Galactic plane. Figure \ref{fig:l_b_400}, shows the $l-b$ diagram resulting from our simulation at $t=2 \Gyr$ ({\it Left panel}). Projected on the plane of the sky, the ring adopts the shape of an asymmetric double-loop between $l\sim-3\grad$ and $\sim +3\grad$, and $b \sim-0\grad.07$ and $\sim +0\grad.1$. On the right panel of figure \ref{fig:l_b_400}, we show the $l-V_{LSR}$ diagram corresponding to the internal region of our Galactic model at $t=2 \Gyr$. The dense, oval structure corresponds to our CMZ since it matches the periodic x2 orbit shown (black line). The $l-V_{LSR}$ diagram found in figure 2 of \citet{armillota2} constructed with NH$_3$ emission data, shows the distribution of the dense and giant molecular clouds that define the CMZ. The NH$_3$ diagram has qualitative similarities with ours since the molecular clouds seem to follow a closed structure in the $l-V_{LSR}$ view. On the other hand, \citet{2017MNRAS.470.1982S} analyzes the three-dimensional structure of the CMZ, studying the molecular gas distribution in the internal region of the Galaxy. The $l-V_{LSR}$ diagram resulting from our simulation, shown in figure \ref{fig:l_b_400}, represents a good qualitative approximation for the morphology of \citet{2017MNRAS.470.1982S} observations. 
While the three-dimensional structure of the CMZ is still debated, the morphological similarities of the ring in the quasi-stationary stage and observations point to a connection between the CMZ and the gaseous ring associated to the x2 orbits that result from the autonomous model.

\section{Gas Flow}
\label{sec:flux}
\begin{figure*}
    \centering
    \includegraphics[width=0.5\columnwidth]{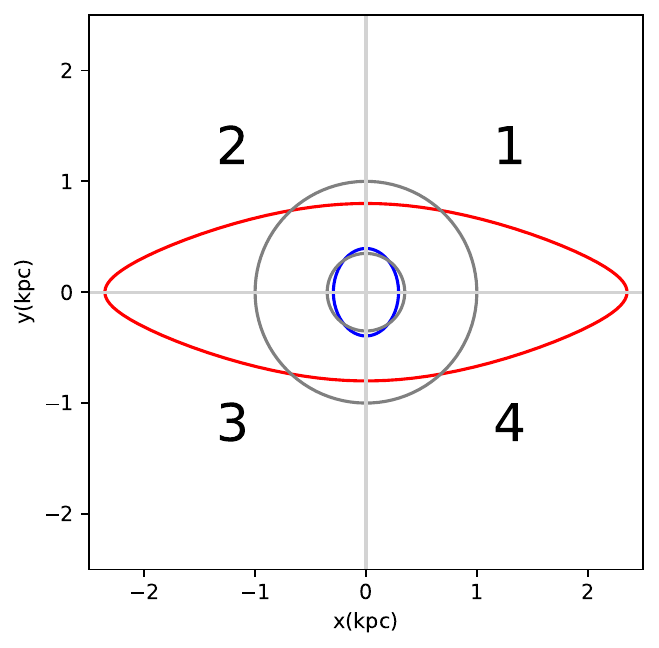}
    \includegraphics[width=0.5\columnwidth]{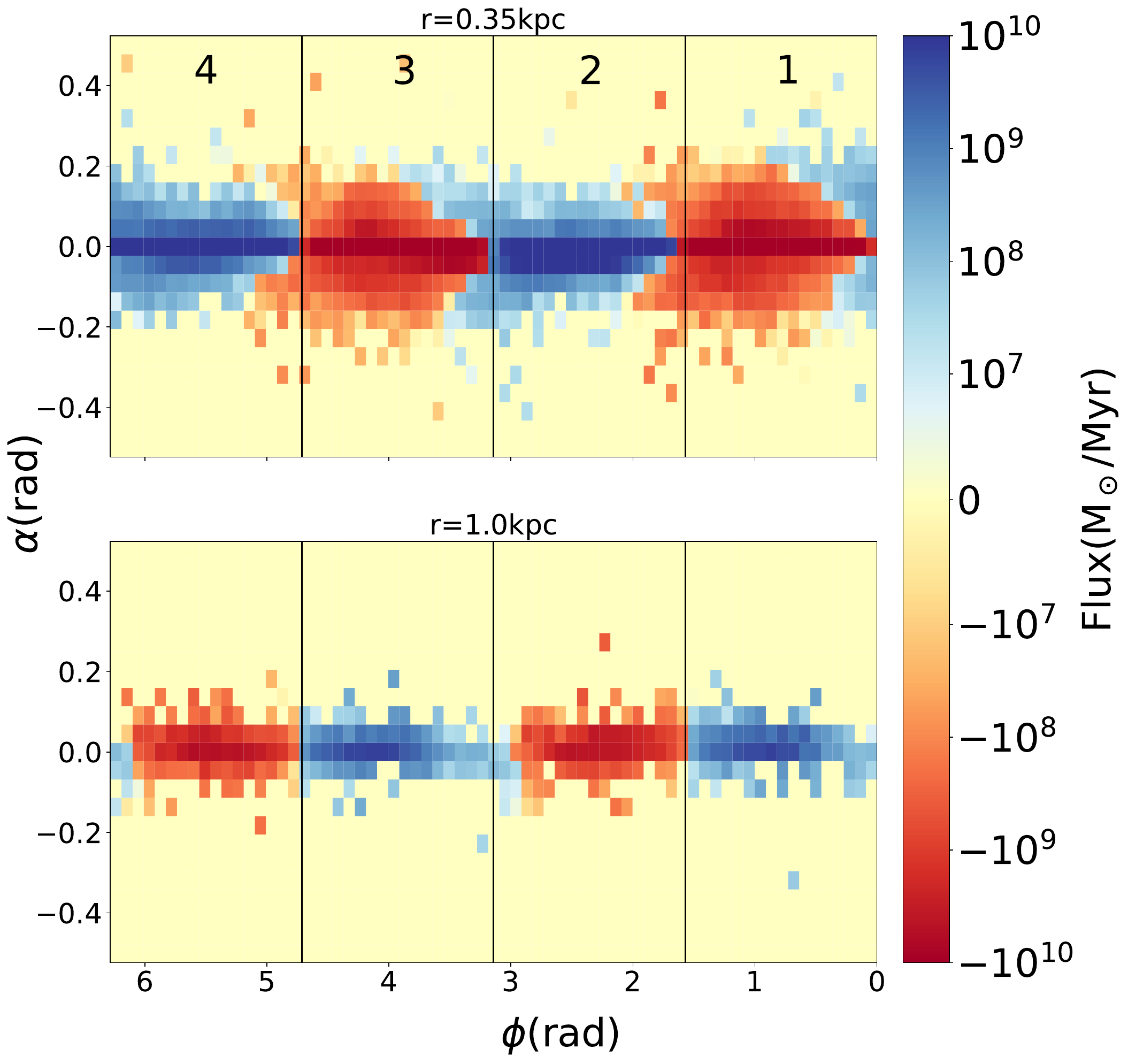}
    \caption{
    ({\it Left}): Division of the galactic plane in quadrants. The blue line shows the location of an x2 periodic orbit corresponding to the position of the inner ring, and the red line shows an x1 periodic orbit from the bar region. The inner and outer gray lines correspond to circles of $0.5$ and $1 \kpc$ radii. Since the rotation is clockwise, a particle (or gas parcel) following the x1 orbit will have positive radial velocities in quadrants 1 and 3, and negative in quadrants 2 and 4. This pattern reverses for a particle following the x2 orbit.
    ({\it Right}): Integrated gas flux measured across $r = 0.35\kpc$ ({\it top}) and $1\kpc$ ({\it bottom}), corresponding to the ring and bar regions, respectively, with positive flux ({\it blue colors}) corresponding to gas flowing towards the external region and negative flux ({\it red}) to gas flowing in towards the inner disk. The vertical lines show the four quadrants of the Galactic plane (labeled on the top plot). The axes correspond to the azimuthal angle in the simulation reference frame ($\phi$; corotating with the galactic bar) and the angle with respect to the galactic plane ($\alpha$).
    }
    \label{fig:orb_flux+stationary_flux}
\end{figure*}
In this section, we discuss the flow of gas during the main phases of the simulation described in section \ref{sec:phases}.
For clarity, we divided the simulation domain into quadrants; the left panel in Figure \ref{fig:orb_flux+stationary_flux} shows this division. Consider a particle (or gas parcel) that follows an x1 orbit (red line). In quadrants 1 and 3, the particle's radial velocity is positive, which implies an outward flux. The opposite happens in quadrants 2 and 4: the particle's negative velocity implies accretion toward the inner galaxy. The situation reverses for particles that follow an x2 orbit (blue line), namely inward flux in quadrants 1 and 3, and outward flux in quadrants 2 and 4. The right panel of Figure \ref{fig:orb_flux+stationary_flux} shows the gaseous mass flux across two constant radius surfaces during the stationary stage of the ring. It can be seen that the gas in the ring (at $r \sim 0.35\kpc$) follows the pattern corresponding to the x2 orbit, while gas in the bar region (at $r \sim 1 \kpc$) tends to follow the x1 orbit.
\begin{figure*}
    \centering
    \includegraphics[width=\textwidth]{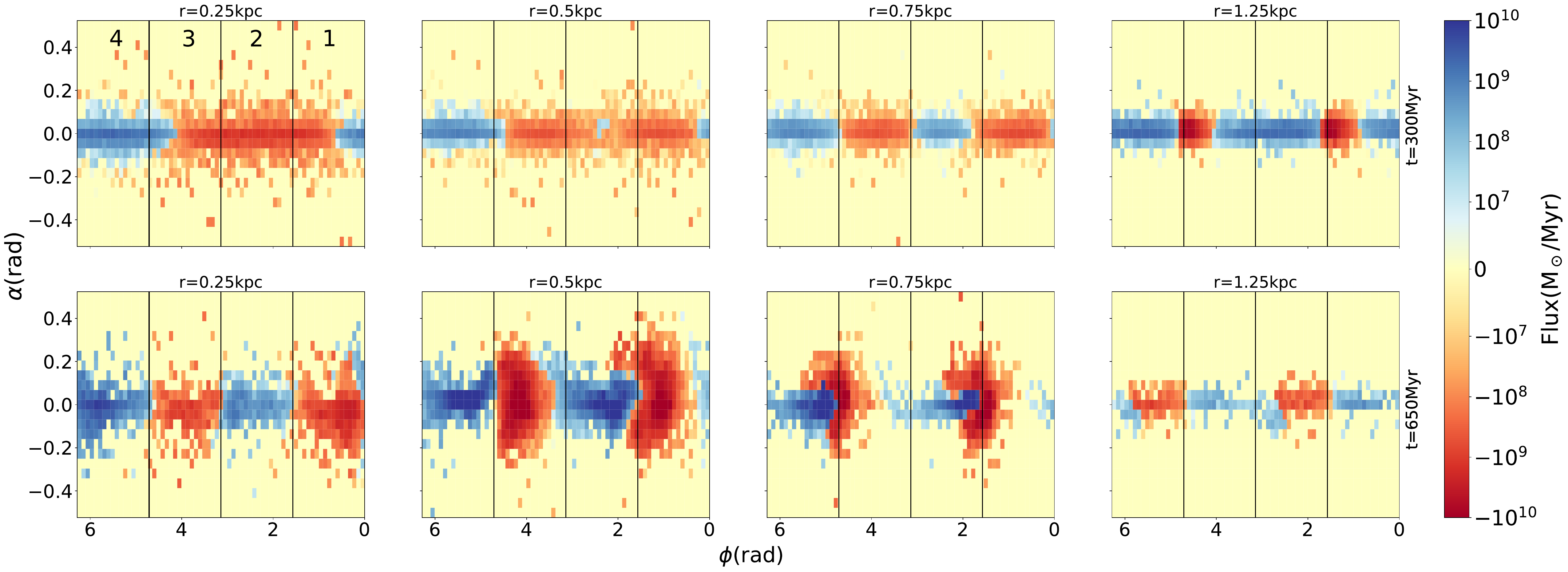}
    \caption{Gas flow during the formation stage across shells of given radius ({\it columns}) at two different times of the simulation ({\it rows}) corresponding to the formation stage of the ring.
    The vertical lines divide each panel in the four quadrants of the galactic plane (fig. \ref{fig:orb_flux+stationary_flux} {\it left}), with each quadrant labeled in the first panel. Positive flux ({\it blue colors}) corresponds to gas flowing out, towards the external regions of the galaxy, while negative flux ({\it red}) corresponds to gas flowing in, towards the inner region.
    The axes correspond to the azimuthal angle in the simulation reference frame ($\phi$; corotating with the galactic bar) and the angle with respect to the galactic plane ($\alpha$).
    }
    \label{fig:flux_all_for_001}
\end{figure*}

As it would be expected, this simple behavior does not happen at earlier evolution times (see sec. \ref{sec:phases}). Figure \ref{fig:flux_all_for_001} shows the integrated gas flux during the formation stage of the ring. At $t = 300 \Myr$ ({\it top panel}), the flux at small radii is dominated by gas that flows inward after losing angular momentum in the spiral shocks visible in Figure \ref{fig:fig_snaps_imp}. On the other hand, at the end of the formation stage at $650\Myr$ ({\it bottom panel}), the inward-outward alternating pattern of the gas flux associated with the x2 orbit is clearly visible, although the inflowing gas has a larger vertical extension than the outflowing gas. Since these extensions are symmetric with respect to the galactic plane, they do not correspond to vertical displacements of the gaseous ring. Instead, they appear to be breathing oscillation modes \citep{WaltersCox91,Widrow_Bonner2015} or vertical motions associated with large scale shocks \citep{MartosCox98,GomezCox04}. These possibilities will be explored in a future contribution.
\section{Star Formation Rate}
\label{sec:sfr}

\begin{figure}
    \centering
    \includegraphics[width=\columnwidth]{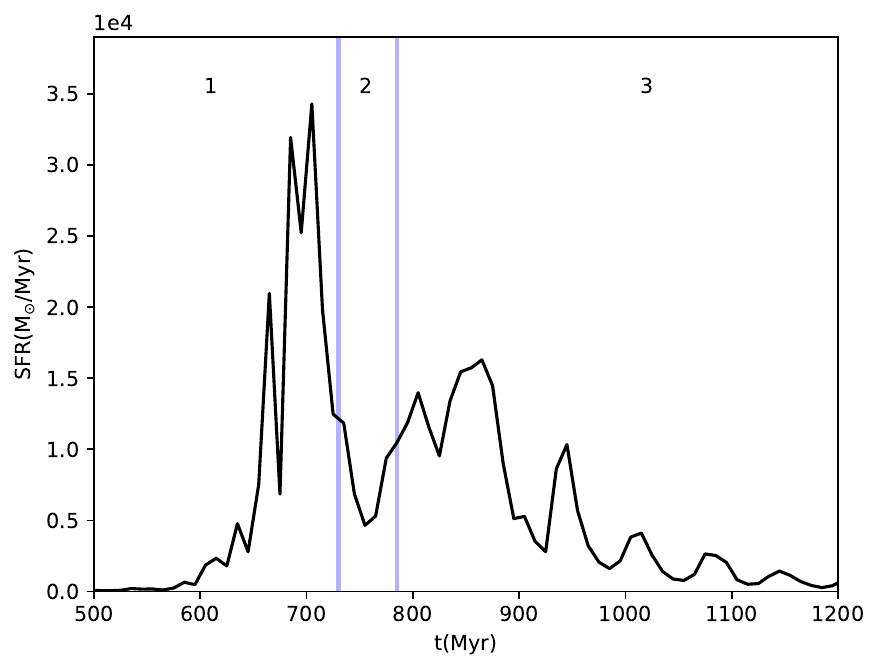}
    \caption{Star formation rate as a function of time.
    Vertical lines show the end of the ring formation stage (region 1), the ring instability (region 2), and the quasi-stationary stages (region 3).
    }
    \label{fig:sfr_001}
\end{figure}
\begin{figure*}
    \centering
    \includegraphics[width=\textwidth]{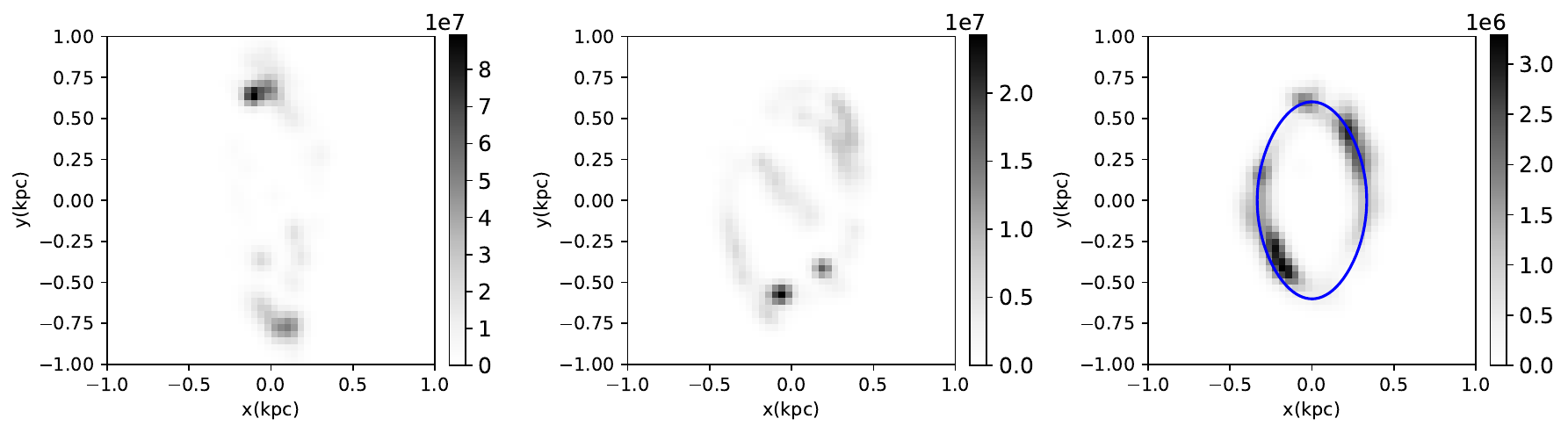}
    \caption{Aggregated star formation sites (in arbitrary units) during the ring formation ($655$ through $730 \Myr$; {\it left}), instability ($730-785 \Myr$; {\it center}), and quasi-stationary stages ($785 - 1560 \Myr$; {\it right}). The grayscale shown is normalized by the time length of each stage. During the formation and quasi-stationary stages, most of the star formation is associated with the orbital apocenters, where gas density is expected to be highest.
    }
    \label{fig:sfnew_001}
\end{figure*}
 
In figure \ref{fig:sfr_001} we show the SFR\footnote{We use the SFR as calculated by the {\sc arepo} code. It uses the sub-grid model presented in \citet{2003MNRAS.339..289S}, which implements star formation and stellar feedback in galaxy simulations.} as a function of time. We observe that most of the star formation activity takes place during the formation stage of the ring, between $t=585 \Myr$ and $t=730\Gyr$. The SFR increases drastically around this stage, reaching a maximum of $\sim 3.5\times 10^{4} \Msun/\Myr$ at $t\sim 700 \Myr$, in agreement with previous determinations of the SFR for the CMZ \citep{Elia_etal2022,Yusef-Zadeh_2009}. It is noticeable that the SFR decreases during the instability stage, a period during which the ring undergoes substantial changes. This might be a consequence of a large amount of velocity shear at this time \citep{seigar2005,colling2018}.
During the quasi-stationary stage, the SFR initially increases again, 
reaching $\sim$ $1.5\times 10^{4} \Msun/\Myr$ at $\sim 860\Myr$. After this time, it decreases until the end of the simulation, with periodic, smaller peaks. After $\sim 1.5\Gyr$, the simulation shows almost null star formation since no gas able to flow into the ring region (see sec. \ref{sec:flux}).

Figure \ref{fig:sfnew_001} shows the locations where star formation takes place, with each panel corresponding to the three identified stages of ring formation.
During the formation stage ({\it left panel}), the ring is more eccentric than in the later stages, and most of the star formation occurs at the apocenters of the orbital distribution, where the dust lanes associated to the galactic bar connect to the ring. This is similar to the simulations reported by \citet{seo1}, which suggest that the star formation occurs mainly at the contact points between the dust lines and the x2 ring (at the apocenters) when the SFR is low, leading to an azimuthal age gradient of young star clusters. 
Also, \citet{2020MNRAS.499.4455T} notice the existence of massive and compact molecular clouds at the apocenters, suggesting that the SFR is extremely high. 
During the instability stage ({\it center panel}), the star formation occurs in a more disorganized fashion, as expected, although the star formation rate is lower than during the formation stage; as mentioned before, this could be a consequence of velocity shear. Finally, during the quasi-stationary stage ({\it right panel}), the ring structure follows a more discernible x2 periodic orbit
, with most of the star formation occurring just \emph{after} the apocenter passage. At this time, the dust lanes are very weak and the gas flow from the bar to the ring is low. So, the location of the star formation events must be linked to the gas orbital dynamics in the x2 configuration and not due to the gas inflow through the dust channels. In this scenario, star formation after the apocenters is not unexpected, since angular momentum conservation implies that the orbital velocity is lowest at the apocenter, and thus the gas density must increase. Although previous work has suggested that the star formation in the CMZ is triggered by tidal fields at the pericenter passage (\citealt{Longmore+13}; see also \citealt{Kruijssen+15}), we do not observe that in our simulation. After the gas density is increased around the orbital apocenter, star formation is triggered and subsequent stages of cloud evolution occur further along the x2 orbit, following the conveyor-belt mechanism described by \citet{Longmore+14}.

\section{Discussion and Conclusions}
\label{conclu}

We performed a hydrodynamic simulation of the gas flow in the inner regions of the Galaxy while a bar potential is being imposed. The simulation develops a ring within $0.5\kpc$ from the galactic center, which we qualitatively associate with the CMZ. The ring development and evolution pass through three main phases, namely formation, instability, and quasi-stationary stages. The formation stage occurs while the bar mass is grown at the expense of the bulge mass. The instability stage occurs after the bar potential is steadied. Since the bar was grown during a period of almost $700\Myr$, we would expect little or no transient structures associated with the bar imposition. Therefore, we do not expect the instability stage of the ring to be a consequence of the numerical setup. Instead, it appears to be a result of the gas being evacuated from the bar region and settling into a stationary orbit.

During the ring evolution, but more clearly in the quasi-stationary stage, the ring can be associated with the x2 periodic orbit family \citep[see also][]{Hatchfield_2021}. Since the gaseous orbits cannot cross themselves, gas parcels can follow periodic stellar orbits only in regions away from resonances \citep{GomezEtal2013}. The three stages of the ring occur in the region between the iILR and the oILR, and so the gas can follow a periodic orbit. Associating the CMZ with the ring in our simulation implies the need for the existence of the x2 orbital family, which in turn 
suggests the existence of a classical bulge in the Galaxy within $\sim 1-2 \kpc$ in addition to the bar \citep{Kunder2020,Queiroz2021}.

During both the instability and quasi-stationary stages, the ring has vertical excursions similar to the ones observed in the CMZ. These vertical oscillations appear to be irregular and it is not obvious if they are associated with off-plane periodic orbital families or with normal oscillating frequencies of the gaseous disk in the central kpc of the Galaxy. It is certainly an intriguing question that will be explored in a future contribution.

The $l-b$ and $l-V_{LSR}$ diagrams corresponding to our simulation show good qualitative agreements to observations. At $t=2 \Gyr$, the $l-b$ diagram shows an infinity shape that has been reported in previous works, while the $l-V_{LSR}$ diagram shows a dense region corresponding to our inner ring that presents an inclined oval-shaped structure. This morphology has also been reported in observations \citep{armillota2,2017MNRAS.470.1982S}.

Most of the star formation in the ring occurs during the formation stage, exactly at the apocenters of the x2 orbit, which correspond to the contact points between the dust channels and the ring structure. 
This case corresponds to the low gas flow rate case discussed in \citet{seo1}. Oddly, the SFR is lower during the instability stage, probably due to the large amount of shear present during this time period. Once the ring settles into the x2 orbit, i.e. during the quasi-stationary stage, the SFR becomes episodic with decreasing amplitude peaks with an $\sim 70 \Myr$ periodicity. Within the ring, star formation occurs mainly after the gas passes its orbital apocenter, where the lower velocity implies higher densities triggering star formation at the first and third quadrants, in contrast with previous models that suggest that star formation should happen near the orbital pericenter \citep{Longmore+13,Kruijssen+15}
or those suggesting that most of the star formation occurs at the apocenters \citep{seo1,2020MNRAS.499.4455T}.
In our simulations, since the preferred sites of star formation occur after the apocenter, we associate it with the dynamics of the x2 orbital configuration and not with a high gas flow at the contact points of the dust lanes. Since the obtained preferred regions of star formation are downstream from the high-density regions at the apocenters, our simulation is consistent with the conveyor-belt scenario \citep{Longmore+14} as applied to the CMZ. 

The details of the 3D distribution of the material in the CMZ are still controversial. Although, in general, it is accepted that the general morphology of the CMZ consists of a ring, finding the actual location of its giant molecular clouds that constitute it is quite challenging from the observational point of view and it is an active topic of study. Since galactic and extragalactic observations suggest that star formation occurs at preferential locations of nuclear rings \citep{2001MNRAS.323..663R, 2006MNRAS.371.1087A, 2008ApJS..174..337M, 2021MNRAS.505.4310C}, it would be useful to establish the positions of the molecular clouds that make up the CMZ to understand the associated star formation. From these clouds, SgrB2, and SgrC present the most active star formation \citep{armillota2}. From this group of molecular clouds, SgrB2 has the better constrained line-of-sight distance, locating it on the foreground of Sgr A* at a galactocentric distance of $R\simeq130\pm 60 \pc$ \citep{2009ApJ...705.1548R}. On the other hand, the distance to SgrC has been poorly constrained \citep{chuard}. The location of SgrB2 in the CMZ seems to be near the apocenter, which favors the hypothesis that star formation is higher in regions close to these points \citep{Hatchfield_2021,2023ASPC..534...83H}. Other works placing the SgrB2 cloud near the apocenter include \citet{2017MNRAS.470.1982S} and \citet{2022MNRAS.516..907S}. It is noticeable that these authors place the SgrC cloud near opposite the SgrB2 cloud. In this contribution, we found that star formation is higher at the apocenters of the x2 orbit during the formation stage, and just after the apocenters during the quasi-stationary state. In the first case, the star formation occurs at the contact points between the dust lines and the ring, while in the second case, the star formation is purely linked to the morphology of the x2 orbits. In both cases, we found that star formation near these points, in agreement with observations that locate the most active star-forming cloud of the CMZ (SgrB2) in one of these regions. The observed distances of these molecular clouds quantitatively disagree with those encountered in the ring in our simulation. However, the locations of active star formation qualitatively agree with observations locating SgrB2 and SgrC in regions close to the apocenters of the observed CMZ.

The link between the x2 orbital family and the gas dynamics (specifically, the sites of star formation) has not been explored in previous studies. Still, it is worth noticing that our models of the CMZ are very simplified and the exploration of the impact of additional physical ingredients, like magnetic fields and stellar feedback, will be required to properly understand the inner region of our Galaxy.

Our simulation aimed to obtain a qualitative model of the physical processes impacting the star formation and dynamics in the CMZ. While the SFR in the simulation is in agreement with observations during the formation stage, the SFR during the quasi-stable stage drops significantly and does not align with observational data. On the other hand, the morphology of the CMZ in the $l-b$ and $V_{LSR}$ diagrams shows some consistency with observational findings reported in the literature. The apparent inconsistency between these two results does not change our conclusions, since the SFR is influenced by physical processes not included in our simulations (magnetic fields and stellar feedback, for example). In this sense, the present study should be considered as a qualitative exploration of the large-scale behavior of the gas associated with the CMZ. A detailed study of the star formation and gas dynamics in the central regions of the Galaxy is still out of reach of current models since there are many unknowns and large observational uncertainties associated with this fascinating structure of the Milky Way.

\section*{Acknowledgements}

We thank C. Benitez for adapting the cooling function to the {\sc arepo} code and an anonymous referee for comments that helped improve this manuscript.
LCV acknowledges support from UNAM-DGAPA fellowship.  GCG acknowledges support from UNAM-PAPIIT IN103822 grant. APV acknowledges support from UNAM-PAPIIT IA103224 grant.

The data underlying this article will be shared on reasonable request to the corresponding author.

\printendnotes

\printbibliography


\appendix

\end{document}